\documentclass[12pt,preprint]{aastex}
\usepackage{amssymb, amsmath, fullpage, graphicx,bm}
\shorttitle{3D simulations of protostellar jets}
\shortauthors{Staff et al.}

\begin{document}
\title{Confronting  3 dimensional time-dependent  jet simulations with HST observations}
\author{Jan. E. Staff}
\affil{Department of Physics and Astronomy, Louisiana State University,
202 Nicholson Hall, Tower Dr., Baton Rouge, LA 70803-4001, 
USA}
\author{Brian P. Niebergal and  Rachid Ouyed}
\affil{Department of Physics and Astronomy, University of Calgary,
2500 University Drive NW, Calgary, Alberta, T2N 1N4 Canada}
\author{Ralph E. Pudritz}
\affil{Department of Physics and Astronomy, McMaster University, Hamilton ON L8S 4M1, Canada
\and Origins Institute, ABB 241, McMaster University, Hamilton ON L8S 4M1, Canada}
\author{Kai Cai}
\affil{Department of Chemistry and Physics,
Purdue University Calumet,
2200 169th Street,
Hammond, IN 46323, USA}

\date{}

\begin{abstract} 
We perform state-of-the-art, 3D, time-dependent simulations of magnetized
disk winds, carried out to simulation scales of 60 Astronomical Units, in
order to confront optical HST observations of protostellar jets. We
``observe'' the optical forbidden line emission produced by shocks within
our simulated jets and compare these with actual observations. Our
simulations reproduce the rich structure of time varying jets, including jet
rotation far from the source, an inner (up to 400 km/s) and outer (less than
100 km/s) component of the jet, and jet widths of up to 20 Astronomical
Units in agreement with observed jets.  These simulations when compared with
the data are able to constrain disk wind models.  In particular, models
featuring a disk magnetic field with a modest radial spatial variation
across the disk are favored.
\end{abstract}

%%%%%%%%%%%%%%%%%%%%%%%
%
% ----- INTRO -----
%
%%%%%%%%%%%%%%%%%%%%%%%
\section{Introduction}
Jets and outflows are important dynamical components of star formation 
and are observed across the entire range of stellar masses, 
from brown dwarfs \citep{whelan05}  to O stars \citep{shepherd05,arce07}.
They appear during the earliest stages of star formation and play a fundamental role in this process by 
removing angular momentum from the gaseous protostellar disks to which they are coupled.  
Jets are known to be associated with all astrophysical systems that have
accretion disks - protostars, black holes of all masses, and compact objects - and may 
even be associated with the formation of giant planets \citep{fendt03}.  However 
the physical conditions and origin of jets can be best studied in the context of protostellar systems  
because of their copious, collisionally excited, optical forbidden line 
emission \citep[eg. review by][]{ray07} and their proximity.

Optical forbidden line emission is the result of collisionally excited
oxygen, sulfur, and nitrogen atoms present in protostellar jets. While the
mechanism(s) responsible for heating the jet and producing the line emission is
still not completely understood, some possible models include radiative heating by X-rays
from the central star \citep{shang02}, magnetic dissipation \citep{moll09}, 
or heating by shocks occurring
throughout the jet volume \citep{ouyed93, shang07}.  Recently, high-resolution
observations of forbidden line emission from jets, obtained with the Hubble
Space Telescope (HST), have been used to directly measure conditions within
protostellar jets \citep{bacciotti99, bacciotti02, ray07}.

A particularly exciting result is the measurement of radial velocity
gradients within jets that provide evidence for jet rotation far away from
the jet's origin \citep{bacciotti02, woitas05, ray07, coffey08}. These results suggest that jet rotation
carries an appreciable amount of angular momentum - of the order of 60 \% of the underlying
accretion disk. This result is crucial because angular momentum must be shed
from the disk in order for material to spiral inwards and build the star.

Theoretical models \citep{bp82, pp92, ferreira97, kk03} and computer 
simulations \citep{ouyed97, machida08} have
shown that an outflowing wind can be launched from magnetized disks. The
magnetic fields that are frozen into the rotating outflow develop a helical
structure resulting in a collimating force, which turns the outflow into a
jet.  Two different hydromagnetic models have been proposed in the literature to explain
the origin of these outflows - disk winds that are launched from
extended regions of magnetized, Keplerian, accretion disks \citep{bp82,
pudritz07}; or X-winds originating from a narrow region on the inner edge of
an accretion disk \citep{shu00}.  An important dynamical distinction between
these models is the amount of angular momentum transported, with disk winds
carrying much more since they originate farther out in the disk. 

The use of HST data to test disk wind models was first demonstrated by
\citet{anderson03}.
Further progress in testing the theories has, until recently, been
hindered by a lack of self-consistent, large scale 3D simulations (matching
HST observations) including the launching and acceleration of the jet from
the vicinity of a protostar and its accretion disk.  The advent of high
performance computing resources has changed this situation.  As an example,
in addition to the work reported here \citet[][ used a simulation box $1/5$
of the length we use]{anderson06} and \citet[][used a simulation box with
the same length as we do]{moll09}.

In our simulations, 
the accretion disk is treated as a boundary condition whose initial conditions
are based on the results of magnetized protostellar disk formation
calculations (see below).  This is a
simplification of the problem. A more satisfying, but also more complicated
approach would require the disk to be part of the simulation. Axisymmetric
simulations of such an initial condition have recently been performed by \citet{murphy10}
and \citet{zanni07} among others. This approach has been limited by the lack
of strong large scale magnetic fields.  Even for this more general coupling
of jet and disk dynamics however,
it is unclear what the appropriate initial state for the disk and jet should  be.  
Moreover, most models assume stationary, 2D, self-similar coupling of jet and
disk solutions, for which there is no general physical justification.

A comprehensive approach to this problem is to start from initial conditions in which
the disk and jet form and evolve together from an initial collapse of 
a magnetized, 3D molecular core 
\citep{banerjee06, machida08, dp09}.  Magnetized collapse produces an  
early outflow that is launched from a small but growing disk.  The strength and the 
geometry of the initial magnetic field still requires further clarification,
\citep[eg.][]{hennebelle08, mellon08}.
\citet{banerjee06} found that $B_z$ is the dominant
component of the magnetic field across the disk and scales with disk radius as $r^{-4/3}$.
The toroidal field component - needed for winds and outflows - is automatically generated in 
the forming rotating disk.
Simulations that also include ambipolar diffusion \citep{dp09} find that the
power law fit to the $B_z$ field varies with disk radius, with the region within 10 AU varying as $r^{-1.67}$ in
one case, and in all cases behaving as $r^{-1.2}$ on larger disk scales.
Jets tap accretion power and therefore simulations need to resolve the inner regions of disks.
The presence of an internal "sink particle" in a disk,
\citep[eg][]{bate95, federrath09} prevents this inner 
region from being resolved, by construction.  Unfortunately this limits the
usefulness of such simulations in tracking
the long term evolution (beyond $10^5$ yr) of the inner few AU of protostellar disks. 

This paper presents simulations with a jet
that propagates out to the equivalent of 60 AU.  We adopt a magnetic field structure on the disks
in accord with the time dependent simulations discussed above. 
Given the difficulties noted above, our approach is
well justified.
A 60 AU outflow scale 
can be directly probed by means of spectroastrometric observations in
forbidden lines such as [OI], [NII], and [SII], using the HST.  As such, we
view our simulated jets in the same lines as observers do in order to
investigate the link between observed jet properties and their actual
underlying dynamics. 
\citet{stute10} followed a similar approach constructing synthetic emission 
line maps from axisymmetric simulations of protostellar jets.

Our results 
show that jets are highly dynamic and riddled with shocks throughout 
their volume.  These heat the jet to produce forbidden line emission.  
Emission maps constructed from our simulations act as an excellent diagnostic for
many aspects of jet collimation, density and temperature structure, and transport
rates of energy and angular momentum from the disk. 
Using these diagnostics, we are able to show that all jet dynamics and
properties are shaped by the magnetic structure, the field strength, and the
mass loading onto the field lines at the base of the flow. We confirm that
jet rotation persists out to at least 60 AU from the source.

%%%%%%%%%%%%%%%%%%%%%%%
% 
% ----- Numerical setup -----
%
%%%%%%%%%%%%%%%%%%%%%%%
\section{Numerical setup}
\label{numericalsection}

Our basic numerical setup consists of a
Keplerian accretion disk as the fixed boundary condition for the
jet \citep{ouyed03}. 
A major addition that we make to earlier 3D work is to incorporate more general models 
for the configuration of the disk magnetic field.  
We present data from two such configurations, whose poloidal magnetic field strength at the disk surface falls off as 
power laws with disk radius as 
\begin{equation}
B_p \propto r_o^{\mu - 1}; \quad \mu = - 0.01, \quad - 0.25.
\end{equation}  
The first model encapsulates initial field structures that can be derived 
from a potential \citep{ouyed1997}
designated OP, while the second models the more steeply decreasing, 
self-similar configuration of Blandford and Payne \citep{bp82} designated BP.  
These models for magnetic structure in the disks reflect those seen in the 
disk formation calculations as noted.
Other than the power law index of the field, the second major 
parameter in our model is the ratio of the thermal and magnetic energy at 
the inner edge of the disk, denoted $\beta_i$.  

Another major extension of previous work is the increase in the spatial
scale of the simulation to 60 AU, compared to the earlier \citep{ouyed03}
box size of less than 2 AU. This is sufficiently large to cover several
pixels of observed jets. We note that the jets shown in our figures have a
much higher angular resolution than can be reached by HST - our images have
resolutions down to 0.015 AU whereas that of the HST is 0.1 arcsec
corresponding to a resolution of about 15 AU \citep{ray07} at the observed
distance of 140 parsec.

We compute the forbidden line emission using the densities and temperatures
of the shocked gas within our simulated jet, allowing us to compare directly
with observations. The shocks produced in the body of our jets are a direct
consequence of the magneto-hydrodynamics. It is well known that flows with
magnetic Mach number $M_A>0.3$ spontaneously generate small, weak, ``eddy
shocklets'' \citep{kida90}. We model the optical emission from the jet by
assuming that any material flowing with $M_A>0.3$ leads to heating by
shocks. This is supported by our previous simulations \citep{ouyed03}, which
show that jets become unstable under these conditions leading to shocks and
jet heating.

We note that our approach is complementary to those which
include the detailed vertical structure of the underlying disk in order to 
model disk winds \citep{ferreira97, kk03}.  This is 
because disk winds and jets - in analogy with the origin of stellar winds - respond to physical conditions 
at their base.  These can be described with a limited set of 
robust boundary conditions.     
Our simulations - when confronted with the HST data - then 
provide highly physical constraints that can in principle constrain more
detailed models that include disk substructure.  

\subsection{Parameters in the model}

Our simulations are technically controlled by the five parameters discussed
in \citet{ouyed1997} and \citet{ouyed03} Those papers did not include the
index of the power law modelling the disk field.  We make many
simplifications of this general scheme. 
Contrary to \citet{ouyed03}, we do not introduce a
toroidal magnetic field in the disk since it is automatically
produced by the outflow. The magnetic field in the disk is
simply an extension of the initial magnetic field in the corona. Similarly,
the plasma $\beta=P_{\rm g}/P_{\rm B}$ at the inner edge of the disk is also
determined by physics - the magneto-rotational instability
\citep[MRI;][]{balbus98}.  The MRI in the disk naturally amplifies weak
fields and saturates at a value $\beta_i \simeq 1$ which we use here. The
Keplerian disk is set up in a similar way to that described in
\citet{ouyed03}, but there are a few differences. Because of the much larger
simulation box, we have extended the outer edge of the disk to $80 r_i$. As
long as this parameter is large, the exact value does not seem to matter.

Contrary to \citet{ouyed03} we do not introduce a turbulent pressure in the
initial disk corona in this
work. The parameter $\delta_i$ used in \citet{ouyed1997} is therefore just
$\delta_i=\frac{\gamma}{\gamma-1}=2.5$, where $\gamma=5/3$ is the polytropic
index. As in \citet{ouyed1997}, the density jump across the
disk surface $\eta_i=100$.

The code uses dimensionless equations, but real units are needed in
order to calculate the forbidden emission lines in the post processing of
the simulation data.
Hence we need the radius of the inner edge of the disk $r_{\rm i}=0.03$ AU,
the mass of the star $M=1M_\odot$, the magnetic field at the inner edge of
the disk $B_{\rm i}=10$ G, the Keplerian velocity at the inner edge of the
disk $v_{\rm k,i}=104{\rm km/s} 
\frac{\sqrt{M/0.5M_\odot}}{\sqrt{r_{\rm i}/0.05 AU}}\approx190 {\rm
km/s}$, the density at the inner edge of the disk $\rho_{\rm i}=9.35\times
10^{-14}\beta_{\rm i}(B_{\rm i}/10 G)^2(r_{\rm i}/0.05 AU)(0.5M_\odot/M){\rm
g cm^{-3}}=5.6\times10^{-14}{\rm g~cm^{-3}}$, and the temperature at the
inner edge of the disk $T_{\rm i}=56.5\times10^4K$. These latter parameters
are only used in the post processing analysis, and therefore do not affect
the dynamics of the jet. They are, however, important for the details of the
results. We have adopted standard values for these parameters (indicated
above) taken from \citet{ouyed1997}.

There are really only two free parameters determining the dynamics in our
simulations, the initial magnetic field configuration (given by $\mu$) and
the mass loading parameter given by the injection velocity \citep[defined as
in][$v_{\rm inj}=0.003$]{ouyed03} from the disk into the corona. The effect
of mass loading was extensively studied in previous 2D simulations
\citep{pudritz07}.  In this work we maintain the mass loading parameter
constant and use two different initial magnetic field configurations.

%%%%%%%%%%%%%%%%%%%%%%%
%
% ----- Methods ---
%
%%%%%%%%%%%%%%%%%%%%%%%
\section{Forbidden line emission}

In these simulations, a polytropic EOS ($P\propto\rho^\gamma$) has been used
($\gamma=5/3$),
and therefore we need not solve the
energy equation.
It was noted in \citet{ouyed03} that no
significant difference was found in the dynamics of simulations containing
the energy equation, and we take this as a justification for using this
simpler approach. We calculated the radiation in the post processing
of the simulation data.

We find the
temperature from the polytropic EOS as:
\begin{equation}
T=\rho^{\gamma-1}
\end{equation}
If the Alfv{\'e}n Mach number ($M_A$) is greater than 0.3, we assume the 
gas to be shocked and the temperature is then \citep{ouyed93}:
\begin{equation}
T=\rho^{\gamma-1} \frac{M_A^2}{\beta} \frac{\gamma-1}{2}
\end{equation}
where $M_A$ is the Alfv{\'e}n Mach number and $\beta=P_g/P_B$ is the plasma
beta.

Given the temperature and assuming standard relative abundances for the
elements in question, the emission from each
cell in the simulation can be found from a well
known relation given by  \citet{haffner99} for [OI]:
\begin{equation}
\epsilon_{\rm OI} = f_\nu dl~exp(-n/n_{\rm crit, OI}) exp(-E_{\rm ij,OI}/kT)
x_e n_{\rm OI} n_{\rm OI} (T/10^4)^{\gamma_{\rm OI}} T^{-0.5}
\Omega(i,j)/\omega_i
\label{emission}
\end{equation}
where $f_\nu$ is the fraction of downward transitions that produce the
emission line, $dl$ is the size of the zone along the line of sight, $n$ is 
the number density, $n_{\rm crit, OI}=1.3\times10^7{\rm cm}^{-3}$, $E_{\rm ij,
OI}=hc/\lambda_{\rm OI}$, $\lambda_{\rm OI}=6300$ \AA, 
$\gamma_{\rm OI}=0.95$ \citep{mendoza83}, k is Boltzmann's
constant, $x_e=0.1$ is the electron fraction \citep{ray07}, $\Omega(i,j)$ is
the collision strength of the transition, and $\omega_i$ is the statistical
weight of the ground level.

It may be instructive to pause for a moment and consider the individual
terms in Eq.~\ref{emission}.
In our simulations, the density $n$ rarely approaches the critical density
for $OI$, $n_{\rm crit,OI}$ so the term $\exp(-n/n_{\rm crit,OI})$ remains
close to unity and does not
play a significant role. The temperature $T$ also rarely approaches $E_{\rm
ij,OI}/k$ meaning that the term $\exp(-E_{\rm ij,OI}/kT)$ does vary a lot
and showing a single peak along the line of sight. Around the peak
this term does not change much, and instead the density squared term 
($n_{\rm OI}^2$) determines the behavior. We note that the
$(T/10^4)^{\gamma_{\rm OI}} T^{-0.5}$ terms behaves similarly to the
$\exp(-E_{\rm ij,OI}/kT)$ term and enhances the temperature dependence.
Since we integrate $\epsilon_{\rm OI}$ along the line of sight, a few
brightly shining cells will completely dominate the emission. 
These dominant cells have high temperature and high density, and as it turns
out they also have high Mach number ($M_A>1$). We chose to include shock
heating for $M_A>0.3$ (see section~\ref{numericalsection}), but as it turns 
out the result is quite insensitive
to this choice and had we chosen to include shock heating for $M_A>1$
instead our results would be similar. The important point to remember is
that we integrate along the line of sight. Even though all the cells with
$0.3>M_A>1$ will have different emission depending on whether shock heating 
is used for
$M_A>0.3$ or $M_A>1$, this is irrelevant for the final result as long as the
line of sight contains dominant cells with $M_A>1$.

We construct forbidden emission lines from our simulations in order to
compare directly with observations.  In order to do the comparison properly,
we should also use the same resolution and integration time as the
observations. Our simulation box cover 8 pixels of HST observations of some
of the nearby jets. If we were to reduce our simulations to only 8 pixels,
most details in our simulations would be lost. We have therefore decided to
maintain this high resolution in our emission line maps to capture more
details and keeping in mind that this resolution is much higher than what
can currently be achieved by observations. The observations have a limiting
magnitude that may be improved in the future, which can reveal some of the
fainter parts of the jet that may otherwise not show up. In this work we
will assume and integration time of ten thousand seconds and a limiting
magnitude of 30 for the resolution that we have.

%%%%%%%%%%%%%%%%%%%%%%%
%
% ----- Results ----
%
%%%%%%%%%%%%%%%%%%%%%%%
\section{Results: diagnostics for jet dynamics}  

We compare the results from the end states of the OP and BP simulations
throughout the rest of the paper. The simulations ran until the front of the
jet reached the end of the grid, which happened after 2300 orbits of the
inner disk (650 days) for the OP simulation and 2550 orbits of the inner
disk (721 days) for the BP simulation. The conversion between code time
units and real time units is given by $t_{\rm orb,id}=0.86\frac{(r_i/0.05
AU)^{3/2}}{\sqrt{M/0.5M_\odot}}$ \citep{ouyed1997}.  Figure \ref{fig1}
presents 3D snapshots of the numerical jet data for these two models.  These
two panels both show the presence of a strong helical field that wraps the
jet, thereby providing the pinch force that collimates the jet towards its
axis.  The bow shock, wherein the external medium is shocked by the jet, is
clearly seen in both cases.

An important difference between these simulations can be discerned.  Whereas 
the BP models appears to have
a density structure that is strongly peaked towards the jet axis, the 
OP model is more bimodal.  
In the OP model there is material at the core of the jet, as well as gas that 
is separated from it.  
The jet creates a cavity outside the narrow inner jet. In the OP case the
material pushed away in this process is collimated by the magnetic field,
becoming the outer jet component. In the BP case
this cavity is wider and opens up faster than in the OP case. The BP magnetic
field (which drops off faster than the OP field) does not collimate this 
material very well, explaining why there is no outer jet component in the BP case.
It turns out that this more extended off-axis component moves at 
much lower velocities than the jet core. 

Both the inner OP jet and the thin BP jet are found to wobble from side to side. 
This is a consequence of a 
finite amplitude instability found for a much more
restricted magnetic geometry by \citet{ouyed03}.  Our present simulations show
that this behavior is quite general - that jets survive nonlinear saturated modes that
give rise to kinks and wobbles and that nevertheless propagate to great distances from
their source.
Movies showing the full time
evolution of these jets can be found at
http://www.phys.lsu.edu/$\sim$astroshare/jstaff/jetmovies/.

Our simulations show that the jet remains stable out to large distances from
the disk. The movies on our web page show that the stable part 
of the jet grows outwards with time.
Closer to the front of the jet the Keplerian velocity profile is lost, which
we attribute to the kink instability. This effect seems to follow behind the
bow shock. The simulations presented in \citet{ouyed03} were on a much
smaller spatial scale (and with a very different initial magnetic field
configuration), so this effect could not be well studied in that work. We
will focus more on jet stability in an upcoming paper.

The integrated [OI] line map for the OP and BP jets are shown in Figure \ref{fig2}.
This map is the sum of all the emission through the whole body of the jet, and 
projected onto the sky. We recall that these maps have resolution of 0.015 AU
in the inner region as compared to 15 AU for HST images.
The core and wider angle structure that we noted previously can be clearly 
seen in the OP jet.
The jet widths can be measured from Figure \ref{fig2}. At about 10 AU
from the disk both jets have a full width of 4-8 AU. Further
away from the disk the bright
inner jet component remains collimated and does not widen much. 
In the OP jet, the dimmer outer component reaches its full width of about 18
AU around 20 AU from the disk.
This is in agreement with 
the observed jet widths \citep{ray07} which are resolved
and found to be about 15 AU to 20 AU close 
to the source. 
The BP jet is less collimated; within our simulation box we find no
outer jet in the BP simulation. The width of the BP jet is therefore just
the width of the thin (inner) jet.
We will discuss jet collimation in more detail in the upcoming paper.
The side boundaries in our simulation box allows for gas to flow out, and a
little bit of gas is indeed leaving through the side boundaries.
A larger simulation box that
minimizes this effect may be preferable as it may reveal an outer jet in the
BP case at a larger radius.

The density and thermal structure in the
OP and BP jets, 
inferred from maps of [OI] emission, are shown in Figure \ref{fig3}.  The figure may be
compared with that of Fig. 6 in \citet{ray07}.  The left and 
right panels show the results for the OP and BP simulations respectively. 
The second row in this figure shows the temperature structure of the jets along two cuts taken parallel to the jet axis - one being along the axis itself and
a second parallel cut displaced from it by 2.5 AU.  
The OP and BP models have a similar temperature structure. We find that in 
the core region of the jet, the temperature is higher than
what was reported in \citet{ray07} but decreasing away from the axis. 
The third row shows the jet 
density inferred by the [OI] diagnostic for
these two simulations.  Again their density
structure is similar and is of the right magnitude, although in the core of
the jet it is higher than what observed jet densities indicate.  At the 
very high spatial resolution shown here,  there
are large fluctuations in the jet density - reflecting the underlying 
noisiness of the jet that is a consequence of its rich shock structure.  
This matches the observed forbidden line emission which also appears to be 
highly variable on similarly small spatial scales \citep{ray07}.  

The fourth row of Fig. \ref{fig3} shows an estimate for the mass loss rates 
that are inferred directly from the simulation data. The 
rate is of the order $~10^{-8}$ M$_{\odot}$ yr$^{-1}$ (adopting
$x_e=0.1$) which is in good agreement with the observed values \citep{ray07}. 
We have taken the jet radius to be 9 AU (OP) or 4.5 AU (BP). The gas outside
of 4.5 AU in the BP jet does not appear to be collimated.
 
Observations of the velocity structure of jets show that jet 
velocities \citep{bacciotti00} can range up to 300-400 km s$^{-1}$. 
We find that OP jets attain slightly higher absolute jet velocities ($\sim440$ km/s) 
than BP models ($\sim400$ km/s).
The lower velocities have a broad spatial
scale, whereas the higher velocity material is much more collimated towards
the axis.
Figure \ref{fig4} very clearly reproduces 
this observed velocity structure.  
We show a series of images of [OI] emission from material that moves in 
the following  velocity channels; $0$-$110$, $110$-$220$, $220$-$330$, and
$330$-$440$ km s$^{-1}$.
The higher velocity material is progressively more collimated and closer to the axis.
The fastest material clearly shows the greatest signs of
a ``wobble'' that is induced by the underlying, kink instability \citep{ouyed03}.
The right panel shows a composite color image of the four velocity channels.

One of the most important tests of hydromagnetic driven wind models for jets 
is that they be observed to rotate \citep{bp82,ouyed97,anderson03}.
 The measured jet rotation and density can be used to measure the angular momentum of the jet, 
and its likely source.  
Using slits placed perpendicular to the jet axis, measured velocity
gradients of 20-30 km s$^{-1}$ can be interpreted as due to jet
rotation \citep{bacciotti02}.

In Figure \ref{fig5} we show the 
magnitude of the expected jet rotation measured by [OI] lines for both the OP and BP
configuration.  
We show that material everywhere in the jet (except in the very front of the
jet) has inherited the
sense of rotation that is imposed by the underlying disk. The shock structure 
induces some local shear gradients on top of this basic pattern. 
In Fig.~\ref{fig6} - we take a "cut" across this forbidden line map (shown
in Fig.~\ref{fig5}) to show the 
measured rotation velocities.  For both OP and BP models, the signature
of Keplerian rotation is unmistakable, and is of the correct magnitude.  
This makes sense since it is ultimately the mass of the central star that sets
the rotation speed of the Keplerian disk, and hence of the jet that arises 
from it.  The rotational signature is more variable in the OP model
because of the wide component. This wide component can also be seen in the
contour plot in Fig.~\ref{fig5}. The BP simulation shows a hint of an
outer rotating component around 10 AU, but due to the limited size of our
simulation box we can not tell if this will evolve into an outer component
as that seen in the OP simulation.
In general, our simulations confirm
that the signature of rotation is robust even in highly varying jets and
matches the predictions of disk wind theory. 

The main difference between the OP and BP simulations is that the OP
simulation collimates most of the wind from the disk into a jet, whereas the
BP simulation only collimates part of it. However, changes in disk
parameters can lead to a range of detailed physical values for jet
quantities.  The
choice of outer radius of the disk has little effect, as not much more mass is lost from a
bigger disk. The width of
the OP jet ($\sim 18$ AU) is a better fit to the observed jet widths. However, by using
$r_i=0.05$ AU (instead of $0.03$ AU) the BP jet becomes $\sim 15$ AU wide,
similar to observed jet widths. Changing $r_i$ also leads to velocities that
are $\sim30\%$ lower than reported here, densities that are $\sim60\%$
higher, and mass fluxes that are $\sim30\%$ higher. In addition, it will
also affect the emission line maps. There are many parameters that can be
changed like this resulting in a range of possible results.  These
parameters are unlikely to be strongly pinned down by observations
in the near future. However, the dynamics remain unchanged by these choices.
 With only two
unconstrained parameters, the mass loading and the magnetic field
configuration, we found that within our simulation box the OP simulation 
collimates the flow much better than the BP simulation.

\section{Discussion and conclusions}

There appear to be two important physical parameters for jets in these
models - the power law index $\mu$ that controls how steeply the magnetic
field falls with radius on the disk at the base of the jet \citep{pudritz06}
and the mass loading.  This paper explores the role of the former - the
latter having been explored in 2D in earlier simulations \citep{op99}. 
Numerical data with the OP model ($\mu = -0.01; \quad B_z \propto r^{-1}$;
which features a more slowly varying poloidal field in contrast with the BP
field $ B_z \propto r^{- 5/4}$) matches observations better.  Since the
magnetic field featured in the X-wind theory falls off even more quickly
with disk radius than the BP configuration, these results suggest that much
more steeply raked magnetic configurations - such as the X-wind
configuration - may have difficulties in predicting the complete structure
of optical emission in jets.

Most of the other parameters that control jet structure and emission in our simulations have values that are established
without any fine tuning.  Emission  
arises from shocked gas that pervades the body of the jet.  Thus, taking all gas with $M_A > 0.3$ as the source of emission
is based on physics.  Secondly, the disk is a region that is susceptible to
magneto-rotational instability \citep{balbus98} whose natural 
saturation results in disk field strengths that are comparable with thermal pressure - hence
$\beta_i \simeq 1$ is 
also the natural setting for this parameter.  We found
that simulations with very high values of $\beta_i$ produced only very low 
velocity jets that do not match the observational data.

Previous efforts have been made to match the HST observations using
stationary, 2D MHD, self-similar disk wind models as a theoretical framework
\citep{dougados04, garcia01}. This approach does not self-consistently
compute wind heating as a consequence of shock dynamics, but assumes it
occurs by ambipolar diffusion in stationary flows.  Such models have been
shown to reproduce the collimation scales of jets if an underlying "warm"
solution is adopted \citep{dougados04}.  There is no direct correspondence
between our simulations and parameters with those of the self-similar
models.  We note however, that the preference for warm solutions - ie ones
which require a disk corona as opposed to a completely cold start for the
MHD disk wind - connects well with our own simulations which have always
proposed that disk coronae exist at the base of disk winds \citep{ouyed97,
op99, ouyed03}.  A second point of contact is the stated importance of the
mass ejection index in the \citet{dougados04} work - which is somewhat
related to our mass loading parameter. Self-similar models, though important
guides, are by construction highly restricted and cannot explore the range
of 3D, time-dependent solutions that are required to understand jet
dynamics.

We conclude that our simulations reproduce many of the observed properties of jets as deduced from their
optical emission lines.  An important result is that the strong signature of jet rotation
observed in our models persists despite rapid time variability and the operation of a 
saturated instability - the jet wobble.  These features do not wash out the overall sense of 
rotation that a jet inherits from its source - the Keplerian accretion disk.  Internal structure in the 
OP model does produce variations in the radial velocity gradients, but overall the interpretation that
jets are rotating is well supported by the observations we make of our 
simulated jets.
Indeed, it is these various instabilities and shocks that make jet emission
possible in the first place. 

Two exciting 
consequences of our work are that jet rotation is not washed out by time dependent
jet evolution, and that our prediction of magnetic field structure on disks could be tested by future ALMA observations.   
Our results open up many fascinating new and testable questions about the nature of jets from disks.  Why are such 
slowly varying disk fields preferred?  Do all jets in protostellar systems have similar magnetic rotors - or does magnetic
field structure in disks evolve with them?   Answers to 
these questions will provide deep insights into the nature of disks and the outflows that they drive.

\begin{figure}
\includegraphics[width=\textwidth]{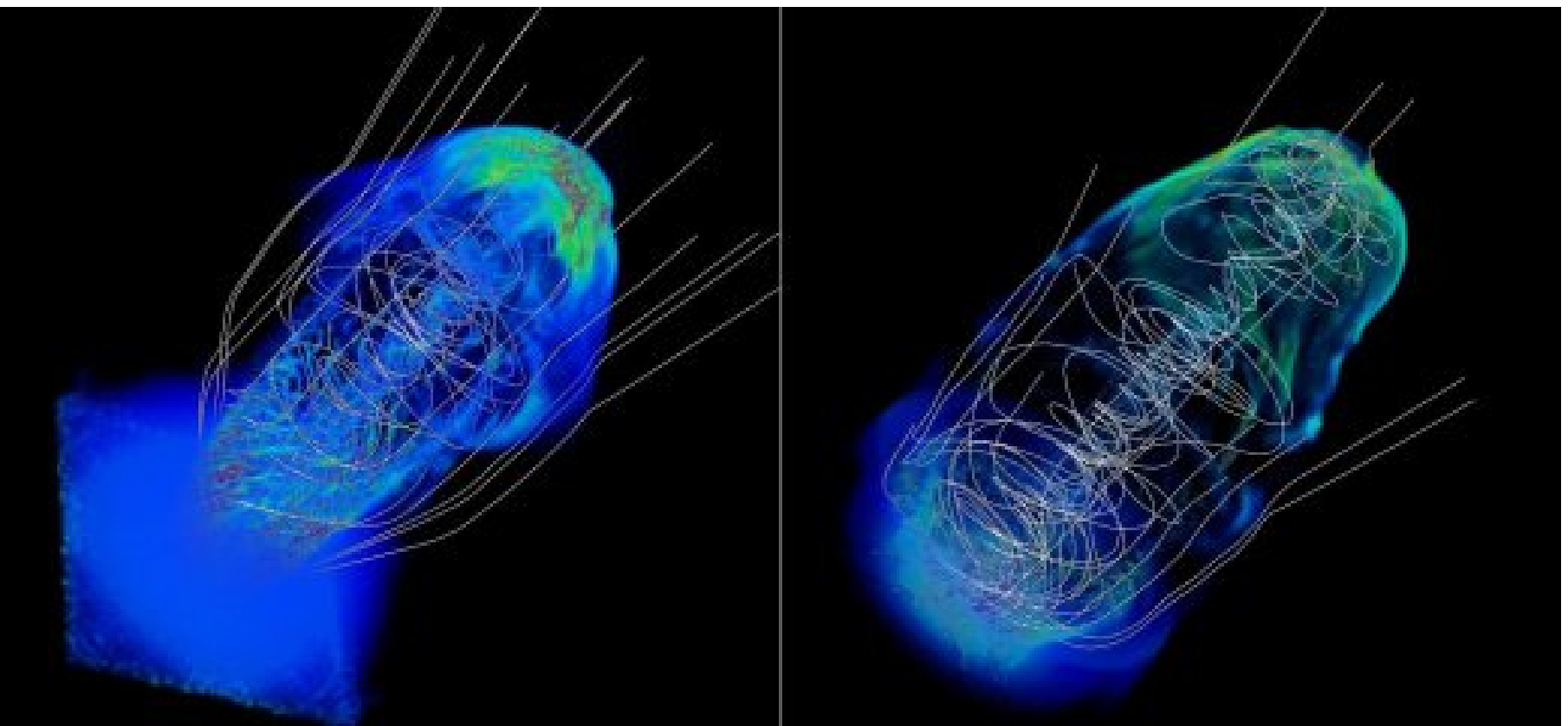}
\caption{{\bf 3D snapshots.}
Density and magnetic field lines for the OP (left panel) and the BP (right
panel) configurations. The snapshots show the rich structure in the jets
when they have reached 60 AU.
The protostar 
and the accretion disk from where the jets are launched is
in the lower left corner of the panels, hidden by the dense coronal material. 
For both jets the bow shock is clearly seen in front of the jet in the 
upper right corners of the figures. Only the OP jet shows two distinct
components, an inner and an outer jet.
\label{fig1}
}
\end{figure}

\begin{figure}
\begin{center}\includegraphics[width=0.8\textwidth]{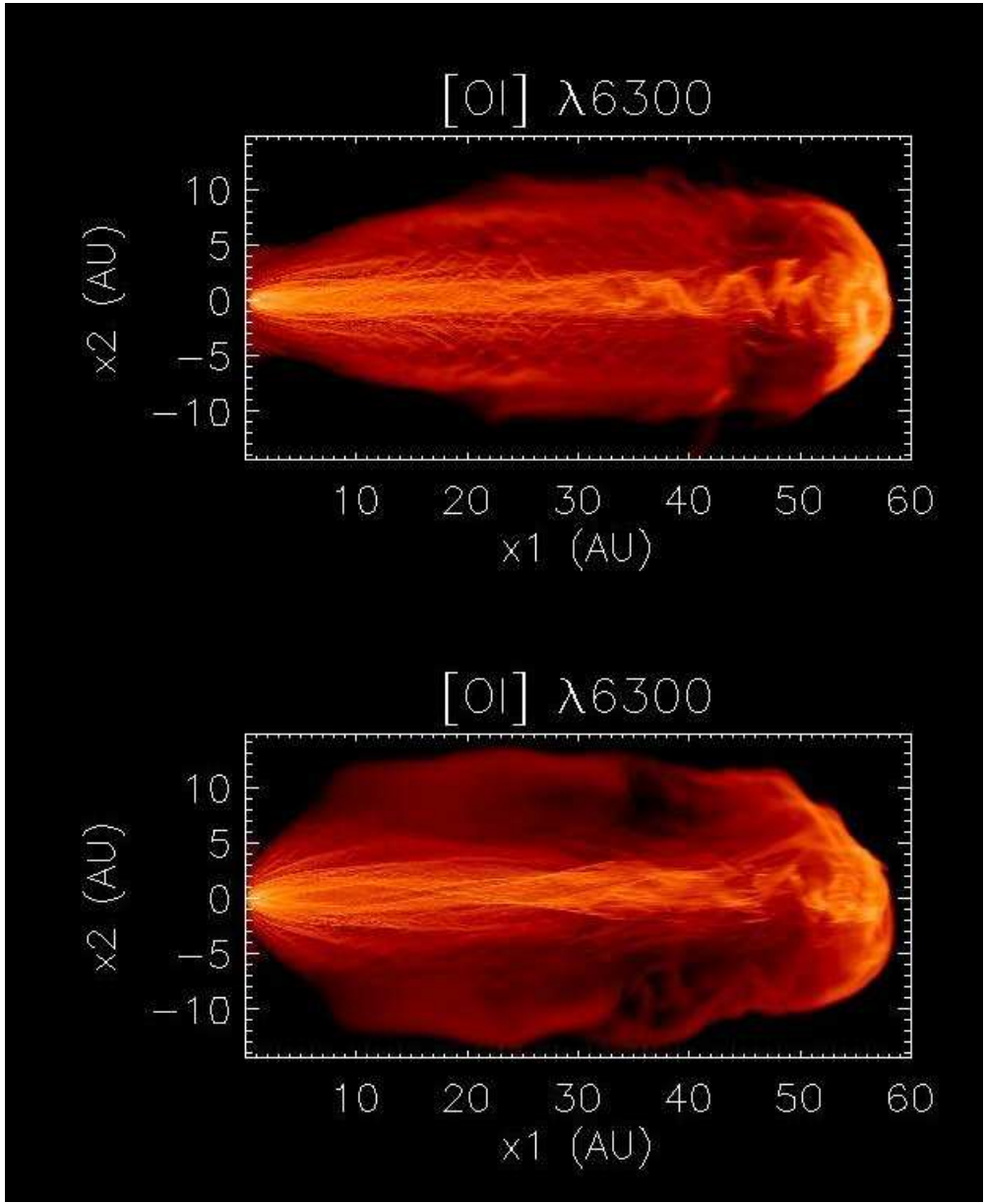}\end{center}
\caption{{\bf [OI] intensity. }
Shown is the [OI] forbidden line intensity map integrated along the line of 
sight, for the OP (upper panel) and BP (lower panel) jet simulations. 
Captured in the OP case is the
spiraling inner jet depicting the kink instability which 
becomes apparent at around 30 AU from the disk. 
The total jet width is found to be up to 18 AU for the OP jet.
The kink mode, although less
apparent, is also present in the BP case. Although a lot of gas is glowing
in the BP simulation, we emphasize that it does not all appear to be collimated.
\label{fig2}
}
\end{figure}

\begin{figure}
\begin{center}\includegraphics[width=1.2\textwidth]{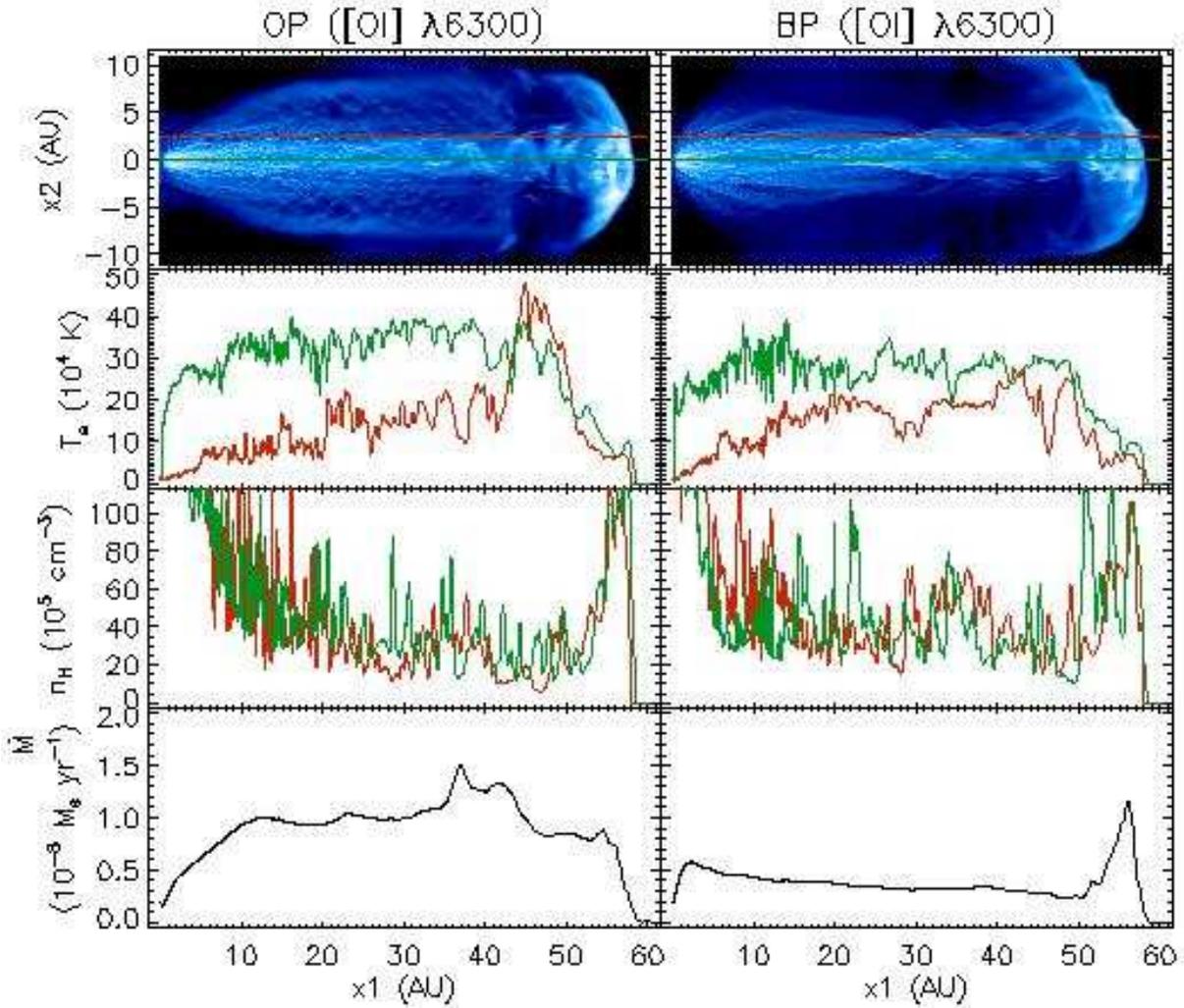}\end{center}
\caption{{\bf Physical quantities along the jet.} The OP configuration is 
shown in the left column and the BP configuration
in the right column. 
The first three rows show from top to bottom: [OI] emission line maps; 
electron temperatures; and number densities. In these three rows the green 
line represents a cut along the center of the
jet whereas the red line is a cut 2.5 AU above the jet axis. 
The bottom row shows an estimate for the mass flux along the jet found
directly from the simulation data. 
\label{fig3}
}
\end{figure}

\begin{figure}
\includegraphics[width=0.99\textwidth]{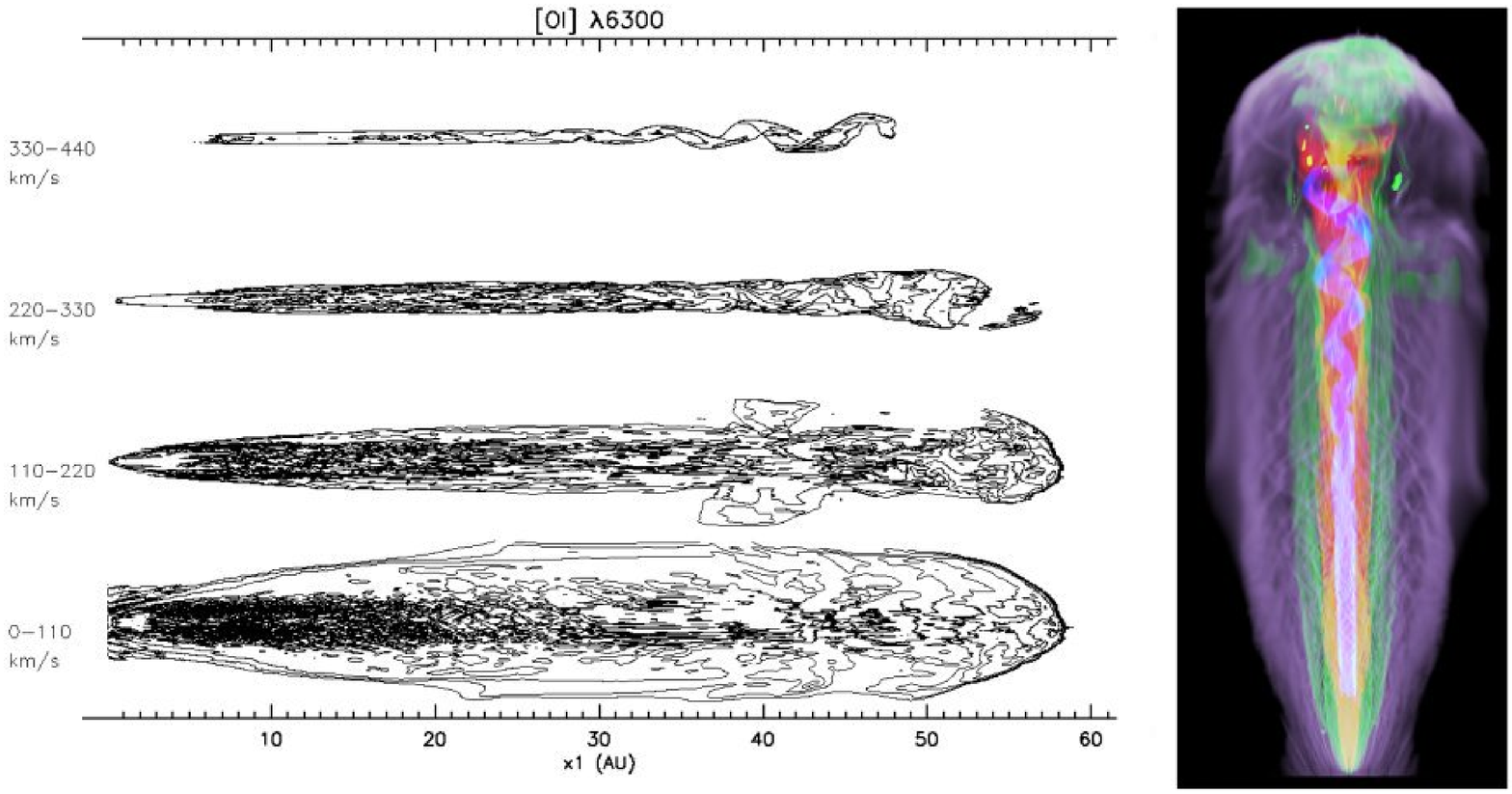}
\includegraphics[width=0.95\textwidth]{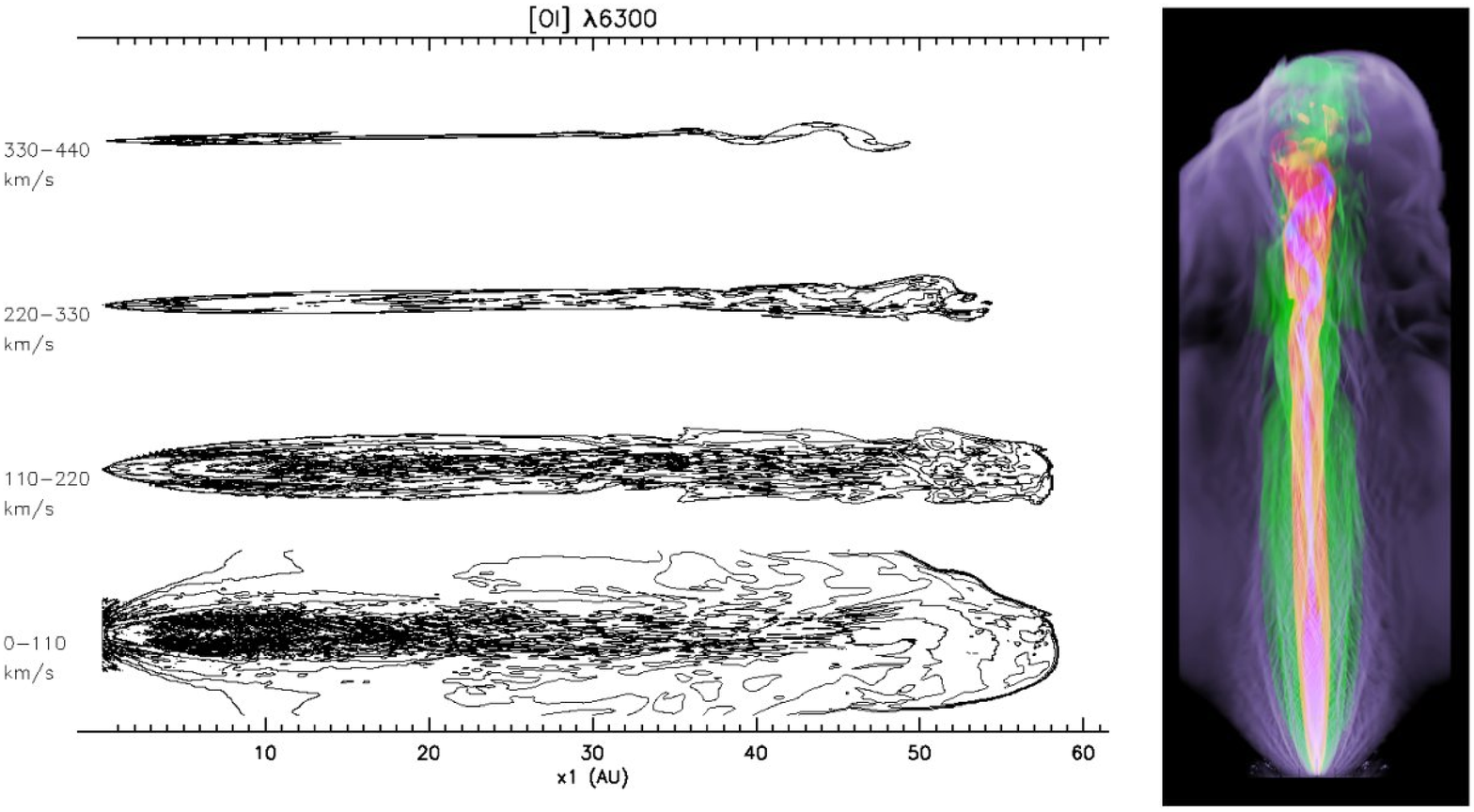}
\caption{{\bf Velocity channel maps.} Shown in the top figure is  
[OI] maps in decreasing (from top to bottom) velocity channels for the OP 
configuration, and similar for the BP configuration in the bottom figure.  
The fastest velocities capture the inner jet and its spiral
structure while the slowest velocities capture the broad outer flow.
In the right panel of each figure is a composite color image of the four velocity channels
(purple (0-110 km/s), green (110-220 km/s),
red (220-330 km/s), and blue (330-440 km/s)).
\label{fig4}}
\end{figure}

\begin{figure}
\begin{center}\includegraphics[width=\textwidth]{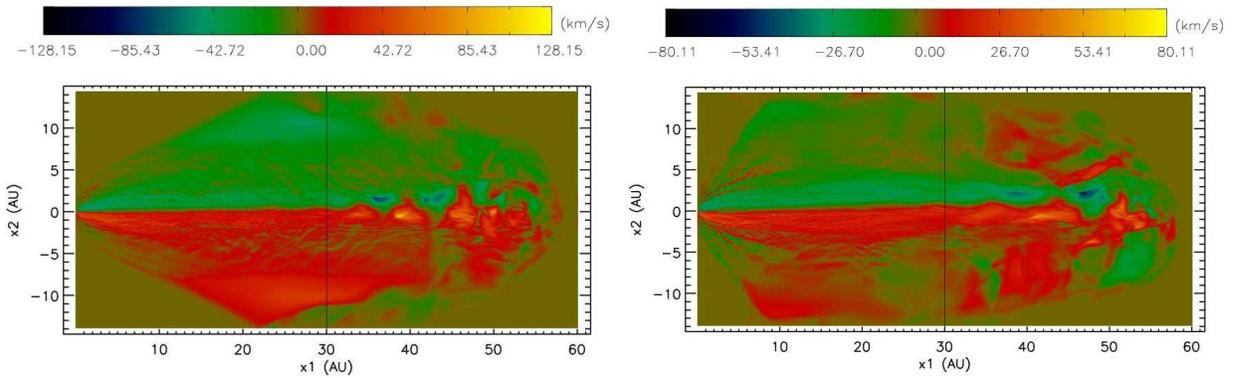}\end{center}
\caption{{\bf Line-of-sight velocity.} The line-of-sight
velocity captures the toroidal velocity from [OI] emission for the OP
configuration (left panel) and BP configuration (right panel). The maximum line-of-sight
velocity is very high ($\sim 100$ km/s) but this high velocity is only reached
a few places in connection with the kink instability. The line-of-sight velocity from
the bulk of the jet is up to about $40$ km/s in both cases.
\label{fig5}}
\end{figure}

\begin{figure}
\begin{center}\includegraphics[width=\textwidth]{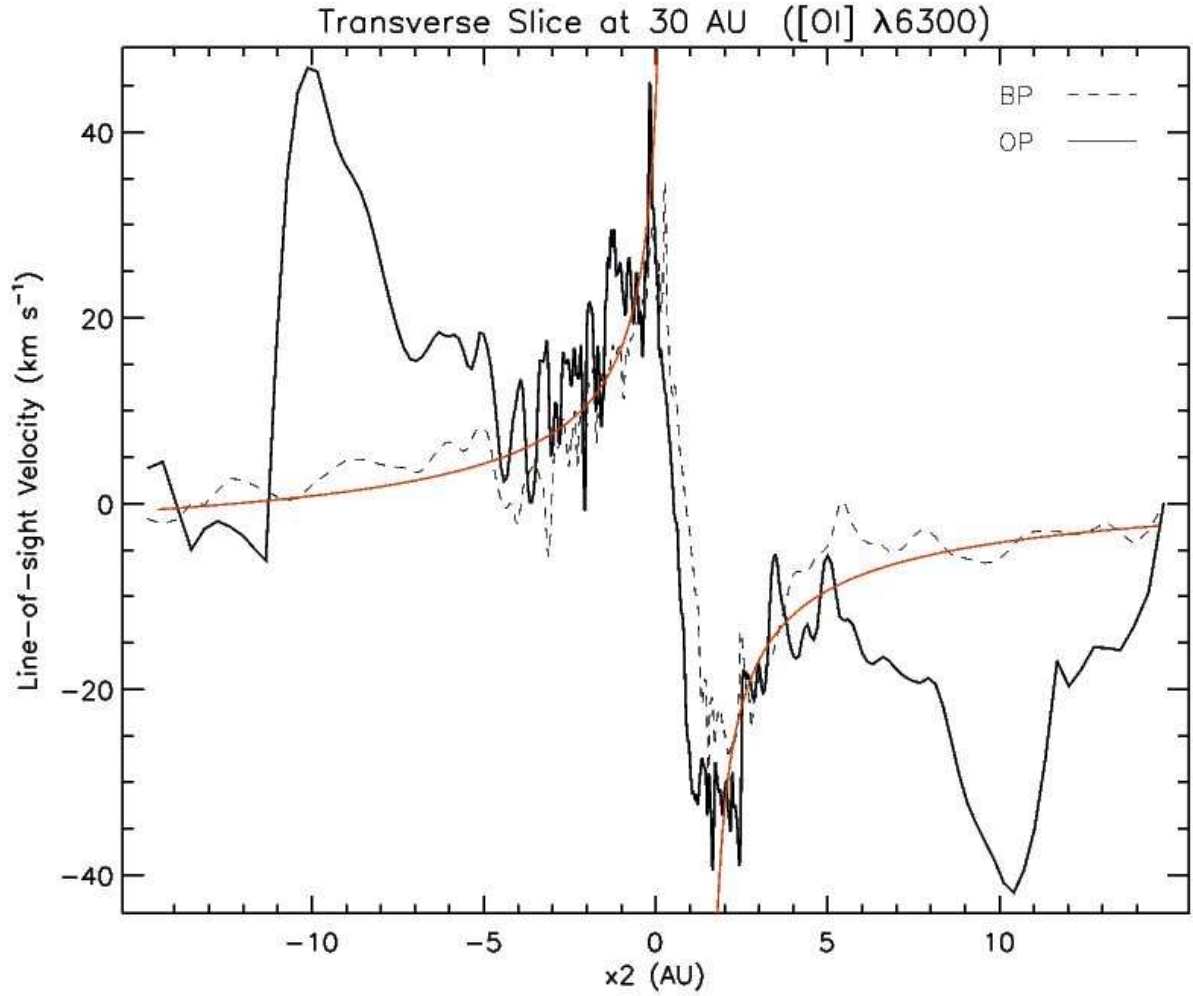}\end{center}
\caption{{\bf Toroidal velocity.}
A slice across the jet taken  
30 AU from the disk (illustrated with a vertical line in Fig.~\ref{fig5}). 
The outer jet component in the OP case (black solid line) starting at about 
5 AU is clearly absent in the BP case (dashed line).
Shown in red is a Keplerian velocity profile that fit OP and BP
line-of-sight toroidal velocity.
The OP rotational profile follows the Keplerian one only until 
about 5 AU, whereafter it deviates because of the outer jet.
\label{fig6}}
\end{figure}

\acknowledgements
We thank J. Ge for help with making Fig.~\ref{fig1}. We also thank the
anonymous referee for a helpful report.
This work has been supported, in part, by grants AST-0708551, PHY-0653369,
and PHY-0326311 from the U.S. National Science Foundation and, in part, by
grant NNX07AG84G from NASA's ATP program. 
RO and REP  are supported by the Natural Sciences and Engineering
Research Council of Canada. 
This work was made possible in part by the facilities of the Shared Hierarchical
Academic Research Computing Network (SHARCNET:www.sharcnet.ca) where we
received a dedicated resource grant of 200,000 CPU hours towards the
project, and in part
by the WestGrid facility. WestGrid computing resources are funded in part
by the Canada Foundation for Innovation, Alberta Innovation and Science, BC
Advanced Education, and the participating research institutions
(WESTGRID:www.westgrid.ca).

%%%%%%%%%%%%%%%%%%%%%%%
%
% Bibliography
%
%%%%%%%%%%%%%%%%%%%%%%%

\end{document}